\documentclass[PAPER, cernpreprint, atlasdraft=false, USenglish, texmf, orcidlogo]{atlasdoc}
\usepackage{atlaspackage}
\usepackage{atlasbiblatex}

\usepackage{atlasphysics}

\usepackage{multirow}
\usepackage{adjustbox}
\usepackage{atlasjournal}
\graphicspath{{logos/}{figures/}}

\usepackage{ANA-STDM-2023-14-PAPER-defs}

\AtlasTitle{Evidence of $ZZ\gamma$ Production with the ATLAS Detector}

\AtlasAbstract{%
This Letter presents the first evidence of the simultaneous production of two $Z$ bosons and one photon with the ATLAS detector at the Large Hadron Collider. The measurement is performed using the full Run-2 dataset, recorded from 2015 to 2018, of proton-proton collisions at a center-of-mass energy of 13 TeV, corresponding to an integrated luminosity of 140 fb$^{-1}$. The fully leptonic final state with four leptons and one photon is analyzed, $pp\rightarrow ZZ\gamma\rightarrow\ell^+\ell^-\ell'^+\ell'^-\gamma$ with $\ell, \ell' = e$ or $\mu$. This final state is measured in a fiducial region where photon final-state radiation is minimized, and the photon has a transverse momentum of $p_{\mathrm{T}}^\gamma > 20$ GeV. Eight events are selected, with a background estimate of $0.92\pm0.15$. This results in an observed (expected) significance of the $ZZ\gamma$ final state of $4.4\sigma$ ($4.0\sigma$). The measured cross-section for $pp\rightarrow ZZ\gamma\rightarrow\ell^+\ell^-\ell'^+\ell'^-\gamma$ in the fiducial region is $\sigma_{ZZ\gamma} = 0.144 _{-0.051}^{+0.065} \text{ (stat.)} _{-0.004}^{+0.008} \text{ (syst.) fb}$, in agreement with the standard model prediction of $\sigma_{\textrm{fid.}}^{\textrm{SM}} = 0.130\pm0.006$ fb.
}

\PreprintIdNumber{CERN-EP-2026-038}

\AtlasJournal{PRL}
\AtlasJournalRef{Phys. Rev. Lett. 137 (2026) 091804}
\AtlasDOI{https://doi.org/10.1103/jngg-1nf3}

\hypersetup{pdftitle={ATLAS document},pdfauthor={The ATLAS Collaboration}}

\begin{document}

\maketitle

Processes with three electroweak gauge bosons in the final state (e.g., $WWW$~\cite{STDM-2019-09,CMS-SMP-19-014}, $WZ\gamma$~\cite{STDM-2019-17,CMS-SMP-22-018}, $W\gamma\gamma$~\cite{STDM-2018-33,CMS-SMP-19-013}, and $Z\gamma\gamma$~\cite{CMS-SMP-19-013,STDM-2021-09})
typically have low cross sections in proton-proton ($pp$) interactions at the Large Hadron Collider (LHC)~\cite{Evans:2008zzb}. Because of the unprecedented integrated luminosity
recorded at the LHC, studies of these processes have recently become feasible.
In this Letter we present evidence of $ZZ\gamma$ production measured with the ATLAS experiment.
The analysis uses the full ATLAS Run-2 dataset of $\sqrt{s} = 13$\,TeV $pp$ collisions corresponding to an integrated luminosity of $140\,\text{fb}^{-1}$. This measurement represents a step forward in probing the gauge structure of the standard model (SM). In the future the $ZZ\gamma$ triple-gauge-boson process can be used to test quartic gauge couplings, including possible deviations arising from new physics~\cite{DEGRANDE201321,PhysRevD.93.093013}, which will benefit from the even larger dataset being recorded in LHC Run-3.

\begin{figure}
\centering
\subfloat[\label{fig:ZZY_feynman_zzy}]{\includegraphics[height=.25\textwidth]{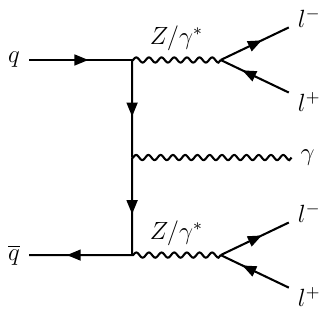}}\hspace*{2.5em}
\subfloat[\label{fig:ZZY_feynman_higgs}]{\includegraphics[height=.25\textwidth]{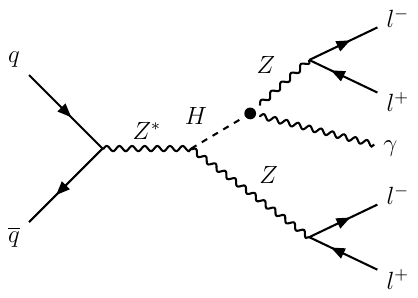}}\hspace*{2.5em}
\subfloat[\label{fig:ZZy_FSR}]{\includegraphics[height=.25\textwidth]{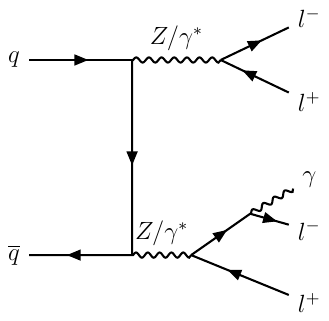}}
\caption{Typical Feynman diagrams producing an $\ell\ell\ell\ell\gamma$ final state for (a) the triboson production, (b) the $Z$-Higgs boson production, and (c) the $ZZ$ production with final-state radiation.}
\end{figure}

In the SM, the $ZZ\gamma$ final state is produced dominantly by Feynman diagrams of the type shown in \Cref{fig:ZZY_feynman_zzy}. A subdominant contribution (1\%) originates through $ZH$ production with the Higgs boson decaying into a $Z$ boson and a photon, as shown in \Cref{fig:ZZY_feynman_higgs}; the effective $HZ\gamma$ coupling is indicated in the diagram. The $ZH$ production yields the $ZZ\gamma$ final state and is taken to be part of the signal definition.
As the Run-2 dataset is limited, an inclusive measurement is performed.
The analysis reconstructs the two $Z$ bosons through their decays into electrons and muons, resulting in four-lepton final states
($e^+ e^- e^+ e^-$, $e^+e^- \mu^+ \mu^-$ or $\mu^+\mu^- \mu^+\mu^-$)
and a photon. The signal definition includes events where a $Z$ boson decays into leptonically-decaying $\tau$ leptons.
The $ZZ$ production with subsequent radiation of
a photon from a final-state lepton [see \Cref{fig:ZZy_FSR}], called
final-state radiation (FSR), can also produce a four-lepton-plus-photon final state.
This contribution is explicitly suppressed by a dedicated rejection criterion, as it is considered to be part of the $ZZ$ and not of the targeted $ZZ\gamma$ production.

\newcommand{\AtlasCoordFootnote}{%
ATLAS uses a right-handed coordinate system with its origin at the nominal interaction point (IP)
in the center of the detector and the \(z\)-axis along the beam pipe.
The \(x\)-axis points from the IP to the center of the LHC ring,
and the \(y\)-axis points upwards.
Polar coordinates \((r,\phi)\) are used in the transverse plane,
\(\phi\) being the azimuthal angle around the \(z\)-axis.
The pseudorapidity is defined in terms of the polar angle \(\theta\) as \(\eta = -\ln \tan(\theta/2)\) and is equal to the rapidity
$ y = \frac{1}{2} \ln \left( \frac{E + p_z}{E - p_z} \right) $ in the relativistic limit.
Angular distance is measured in units of \(\Delta R \equiv \sqrt{(\Delta \eta)^{2} + (\Delta\phi)^{2}}\).}

The ATLAS experiment~\cite{PERF-2007-01} at the LHC is a multipurpose particle detector with a cylindrical geometry.\footnote{\AtlasCoordFootnote}
It consists of an inner tracking detector surrounded by a thin superconducting solenoid providing a \qty{2}{\tesla} axial magnetic field, electromagnetic and hadronic calorimeters, and a muon spectrometer
based on
three large superconducting air-core toroidal magnets with eight coils each.
The luminosity is measured mainly by the LUCID-2~\cite{LUCID2} detector that is located close to the beam pipe.
A two-level trigger system is used to select events~\cite{TRIG-2016-01}.
A software suite~\cite{SOFT-2022-02} is used in data simulation, in the reconstruction and analysis of real and simulated data, in detector operations, and in the trigger and data acquisition systems of the experiment.

The data in this analysis were collected between 2015 and 2018 in $pp$ collisions at $\sqrt{s} = 13$\,TeV. Only events recorded with a fully operational detector and stable beams are used~\cite{DAPR-2018-01}. The resulting integrated luminosity is $140$\,fb$^{-1}$ with an uncertainty of $0.83\%$~\cite{DAPR-2021-01}.
Signal and background processes were simulated with several Monte Carlo (MC) event generators, and the ATLAS detector response was modeled with \textsc{Geant4}~\cite{Agostinelli:2002hh,SOFT-2010-01}.
Data and simulated events are processed with the same reconstruction and analysis software.

The $ZZ\gamma$ signal is simulated using a \textsc{Sherpa} 2.2.11 $4\ell\gamma$ sample~\cite{Bothmann:2019yzt}. The sample was generated at
next to leading order in perturbative QCD
with zero additional partons and at leading order for up to two
additional partons; the matrix elements were merged with the \textsc{Sherpa} parton shower~\cite{Schumann:2007mg} according to the CKKW procedure~\cite{Hoeche:2009rj}.
The \textsc{NNPDF3.0} next-to-next-to-leading-order parton-distribution-function (PDF) sets are used~\cite{Ball:2014uwa}. The sample includes both on- and off-shell $Z$ boson contributions, next-to-leading-order electroweak corrections, and FSR photons.
As the Higgs boson contribution is not included in that sample, an additional signal sample for $ZH\rightarrow ZZ\gamma$ was generated with \textsc{Powheg-Box}~\cite{Nason:2004rx,Frixione:2007vw,Alioli:2010xd} interfaced to \textsc{Pythia} 8.210~\cite{Sjostrand:2014zea}. The interference between the two processes is neglected, given the small width of the Higgs boson.

The signal region (SR) is defined by the presence of two opposite-sign same-flavor (OSSF) lepton pairs (electron or muon) and one photon with high transverse momentum (\pt); no additional requirements are placed on other objects.

Photon and electron candidates are reconstructed from energy clusters in the electromagnetic calorimeter and combined with information from particle tracks in the inner detector, as described in Ref.~\cite{EGAM-2018-01}. The selected electron (photon) energy clusters must lie within $\vert \eta \vert < 2.47 \ (\vert \eta \vert < 2.37)$, with the transition region between the barrel and the end-cap calorimeters excluded ($1.37 < \vert \eta \vert < 1.52 $). Muon candidates are reconstructed by combining tracks from the muon spectrometer with inner detector tracks, according to Ref.~\cite{PERF-2015-10}, and are required to fulfill $\vert \eta \vert < 2.5$. All lepton candidates must originate from the primary vertex, which is defined as the vertex with at least two tracks that has the highest sum of squared transverse momenta of its associated tracks. The significance of each electron (muon) track's transverse impact parameter, $d_0$, must satisfy $\vert d_0\vert/\sigma_{d_0} < 5\ (3)$ and the longitudinal impact parameter, $z_0$, must fulfill $\vert z_0\ \text{sin} (\theta)\vert < 0.5\,$mm~\cite{ATL-PHYS-PUB-2015-026}. Electron (muon) candidates are required to pass the \textit{Medium} identification criteria and the \textit{Loose} (\textit{PflowLoose}) isolation criteria given in Refs.~\cite{EGAM-2021-01,MUON-2018-03}, and have $p_{\mathrm{T}} > 10\,$GeV. One or two of the lepton candidates must have fired an unprescaled single or dilepton trigger~\cite{TRIG-2018-05,TRIG-2018-01}. Additionally, photon candidates must have $p_{\mathrm{T}}^\gamma > 20\,$GeV, satisfy the \textit{Loose} isolation and \textit{Tight} identification criteria of Ref.~\cite{EGAM-2021-01}, and be separated from lepton candidates. If a lepton-photon pair fulfills $\Delta R (\ell,\gamma) < 0.4$, the photon is discarded. Likewise, electrons found within $\Delta R (\mu,e) < 0.2$ of a muon candidate are removed from the event.

The photon candidate with the highest transverse momentum, \pTy, is selected. The \pt of the leading and subleading leptons must be greater than 30 and 25~\GeV, respectively. The two OSSF lepton pairs with invariant masses ($m_{\ell\ell,1}, m_{\ell\ell,2}$) of at least $40\,$GeV and closest to the $Z$ boson mass $m_Z$~\cite{ParticleDataGroup:2024cfk} are chosen, minimizing ($\vert m_{\ell\ell,1} - m_Z\vert + \vert m_{\ell\ell,2} - m_Z\vert$).
If an event contains an FSR photon that carries a sizable fraction of the $Z$-boson decay energy, the pairing can become ambiguous. This ambiguity is resolved by imposing additional requirements. To reject FSR events, the sum of the three-body invariant mass $m_{\ell\ell\gamma}$ and $m_{\ell\ell}$ of each of the OSSF lepton pairs must exceed twice the $Z$ boson mass, $(m_{\ell\ell\gamma} + m_{\ell\ell}) > 2 m_Z$. This is motivated by the fact that, in case of an FSR photon, the three-body invariant mass $m_{\ell\ell\gamma}$ is expected to be near $m_Z$, while $m_{\ell\ell}$ of the same lepton pair tends to be smaller than $m_Z$~\cite{STDM-2018-04,STDM-2021-09}.

The \ZZy\ production cross section is measured in a fiducial phase-space region defined at particle level with kinematic requirements that closely match those defining the detector level SR. The four-momenta of photons found within a cone of size $\Delta R = 0.1$ around each electron or muon are added to its four-momentum.
The \pt and invariant mass requirements on the four signal leptons and the photon are identical to the ones in the SR. The pseudorapidity requirement $|\eta^\ell| < 2.5$ ($|\eta^\gamma| < 2.37$) is imposed on leptons (photons). Photon isolation is imposed by demanding that the scalar sum of the \pt of all stable particles ($c\tau>10$~mm) within a cone of size $\Delta R = 0.2$ around the photon is less than 7\% of the photon \pTy.

The dominant background arises from nonprompt photons, i.e., photons coming from hadron decays, and accounts for roughly 75\% of the total background yield. A second source is nonprompt leptons---leptons from hadron decays or misidentified jets and photons---contributing around 15\%. The remaining 10\% are due to pileup; additional $pp$ collisions that occur in the same or in neighboring bunch crossings. Other background contributions were evaluated and found to be negligible.
Nonprompt photons are mainly produced from $\pi^0$ decays within jets and are typically surrounded by hadronic activity. The photon isolation requirements suppress this background significantly. The background that remains is estimated with data-driven techniques. Because of the small data sample, a nonprompt photon ratio method is applied. The underlying idea is that the fraction of jets misidentified as nonprompt photons will be approximately independent of the number of $Z$ bosons in the event. This is justified by the fact that the photons arise from the hadronization of the jet, a process that does not depend on the electroweak part of the event. Under this assumption one obtains the following relation:
\begin{equation}
\frac{N_{ZZ\gamma,\,\gamma_{\text{nonprompt}}}}{N_{ZZ+\text{jets}}}=\frac{N_{Z\gamma,\,\gamma_{\text{nonprompt}}}}{N_{Z+\text{jets}}} \ \Rightarrow\ N_{ZZ\gamma,\,\gamma_{\text{nonprompt}}}=N_{Z\gamma,\,\gamma_{\text{nonprompt}}} \frac{N_{ZZ+\text{jets}}}{N_{Z+\text{jets}}}.
\label{eq:fake_ratio}
\end{equation}
The number of events with a nonprompt photon in the SR is denoted by $N_{ZZ\gamma,\,\gamma_{\text{nonprompt}}}$, while the number of events with a nonprompt photon in the $Z\gamma$ control region (CR) is $N_{Z\gamma,\,\gamma_{\text{nonprompt}}}$. The $Z\gamma$ CR is defined similarly to the SR but without the second OSSF lepton pair. Additionally, $N_{Z+\text{jets}}$ and $N_{ZZ+\text{jets}}$ are the number of events with at least one jet and one or two $Z$ bosons, respectively.

For Eq.~\ref{eq:fake_ratio} to be a good approximation, the phase-space requirements of all selections must be as similar as possible. The single-$Z$ events satisfy the same criteria as the leading $Z$ boson in the di-$Z$ selection. Using particle-flow objects~\cite{PERF-2015-09} as input, jets are reconstructed via the anti-$k_T$ algorithm~\cite{Cacciari:2008gp, Fastjet} with $R=0.4$, and the highest \pT\ jet is required to pass the same \pT\ requirement as the photon, $p_{\mathrm{T}} > 20\,$GeV. To account for the jet width, jets within $|\eta|<2.6$ are allowed, and the transition-region veto is not applied.

The number of nonprompt photons in the $Z\gamma$ CR is estimated via a data-based purity method. The purity $P$ is defined as the fraction of $Z\gamma$ prompt photon events relative to the total number of $Z\gamma$ prompt and nonprompt photon events. Its values range from 0.75 to 0.98 increasing with \pTy\,. The purity is taken from a two-dimensional sideband method used to estimate the $Z+$jets background in the previous $Z\gamma$ analysis~\cite{STDM-2018-04}. For the data-driven purity method, the total number of $Z\gamma$ events---with prompt and nonprompt photons---in the $Z\gamma$ CR is needed. The $t\bar{t}\gamma$, $WZ$, and $ZZ$ background contributions (3.3\%) are estimated with MC and subtracted from data in the $Z\gamma$ CR. The remaining events are multiplied by $(1-P)$ to obtain the number of nonprompt photon events in the $Z\gamma$ CR. As a cross-check, a MC-based purity estimation is performed with the same \textsc{Sherpa 2.2.4} $Z\gamma$ sample as used in Ref.~\cite{STDM-2018-04}. In this case the $Z\gamma$ MC yield in the $Z\gamma$ CR is scaled by $(1-P)/P$ to give the nonprompt photon component. The results of the two methods agree within uncertainties.

The ratio of $ZZ+$jets and $Z+$jets events is taken from data. The ratio shows a clear dependence on the transverse momentum of the jet but no dependence on other kinematic variables (e.g., $\eta$, $\phi$). Multiplying the estimate of the number of nonprompt photon events in the $Z\gamma$ CR by the ratio of $ZZ+$jets and $Z+$jets events as given by Eq.~\ref{eq:fake_ratio} in bins of $p_{\mathrm{T}}$ yields $N_{ZZ\gamma, \gamma_{\text{nonprompt}}} = 0.71 \pm 0.02 \text{ (stat.)} \pm 0.13 \text{ (syst.)}$ in the SR.

The systematic uncertainty of $N_{ZZ\gamma, \gamma_{\text{nonprompt}}}$ comes from the uncertainty in the purity ($14\%$)~\cite{STDM-2018-04}, the uncertainty in the jet component differences from the ratio of quark- and gluon-induced jets in $ZZ+$jets and $Z+$jets events derived from MC samples ($10\%$), the difference with respect to the MC-based purity estimate to account for phase-space differences to the $Z\gamma$ analysis ($7\%$), and the luminosity uncertainty ($<1\%$). The method is validated by replacing one $Z$ boson with a $W$ boson and shows good agreement with the nonprompt photon estimate in the $WZ\gamma$ analysis~\cite{STDM-2019-17}.

The second source of background comes from nonprompt leptons and is estimated with the matrix method~\cite{EGAM-2019-01}. The method uses the efficiencies of the prompt and nonprompt leptons to pass the tight SR selection relative to a looser baseline selection. Baseline electrons have no isolation requirements and a \textit{Loose} identification~\cite{EGAM-2021-01}. Baseline muons have a looser transverse impact parameter requirement, $\vert d_0\vert/\sigma_{d_0} < 10$, and no isolation requirement. All other selection criteria are the same as in the SR. A nonprompt lepton-enriched $Z+1\ell$ CR is defined, where exactly three leptons are required and no photon is selected. Two of the three leptons must fulfill the $Z$ boson requirements of the SR. Prompt lepton contributions from $WZ$ and $ZZ$ processes, estimated with MC simulations, are subtracted from the data in the $Z+1\ell$ CR. The prompt-lepton efficiency is obtained with the \textsc{Sherpa} 2.2.11 $4\ell\gamma$ signal sample; the nonprompt-lepton efficiency is measured in the $Z+1\ell$ CR. Because of the limited number of events, the nonprompt lepton contribution is estimated in a $ZZ$ CR---the SR without any photon requirements---and extrapolated to the SR. A $30\%$ modeling uncertainty is applied to account for the different lepton \pT spectra in the $ZZ$ CR and the SR. The background estimate in the SR is $N_{ZZ\gamma}^{\text{nonprompt }\ell} = 0.12 \pm 0.02 \text{ (stat.)} \pm 0.08 \text{ (syst.)}$. The two largest systematic uncertainties are the uncertainty on the prompt lepton contributions estimated from MC simulations in the $Z+1\ell$ CR and from the modeling uncertainty.

The final source of background, the pileup background, consists of the combination of a $ZZ$ event with a photon produced in a different $pp$ collision, entering the sample due to the poor reconstruction of the photon's point of origin. It is estimated with a method similar to the one used in Ref.~\cite{STDM-2018-04}: particle level $pp\rightarrow ZZ\rightarrow4\ell$ events are overlaid with photons from inclusive $\gamma+$jet production. The events in the fiducial region are selected and a correction factor is applied to derive the pileup contribution in the SR. The correction factor accounts for limited detector resolution and reconstruction inefficiencies and is derived with the $4\ell\gamma$ sample. This results in a pileup contribution of $N_{\text{PU}}= 0.089 \pm 0.002 \text{ (stat.)} \pm 0.022 \text{ (syst.)}$.

To extract the \ZZy\ signal event contribution in the SR, a profile-likelihood-ratio test statistic is constructed in the asymptotic approximation~\cite{Cowan:2010js} using the data and backgrounds estimated as above.
This determines a signal-strength parameter $\mu_{\ZZy}$, which measures the signal contribution relative to the SM expectation.
All leptonic final states are fitted simultaneously, using the estimated background contributions, the observed numbers of events, and the prefit yield of the expected signal.

Systematic uncertainties affecting the cross section measurement include contributions from electron trigger, reconstruction, identification~\cite{EGAM-2021-01} and isolation requirements (0.6\%), photon identification and isolation requirements (1.6\%), electron-photon resolution and energy scale~\cite{EGAM-2021-02} (0.5\%), muon trigger, reconstruction, identification~\cite{MUON-2018-03} and isolation requirements, and momentum resolution and scale~\cite{MUON-2018-03} (0.9\%), modeling of the pileup in the MC samples (1.4\%), luminosity (0.9\%), and theoretical modeling of \ZZy\ events (1.4\%). The theoretical modeling is estimated by varying the renormalization and factorization scales, as well as the PDFs and \alphas, according to Ref.~\cite{ATL-PHYS-PUB-2016-002,Butterworth:2015oua}. In addition, the uncertainty in the predicted SM signal yield due to a limited MC sample size (1.4\%) and the background uncertainties (2.3\%) contribute to the systematic uncertainty in the measured cross section in the fiducial region, $\sigma_{\ZZy}$.
The total systematic uncertainty is 4.0\%.
The measurement is statistics limited, with about 40\% statistical uncertainty.

\begin{figure}
\centering
\subfloat[]{\includegraphics[width=0.49\textwidth]{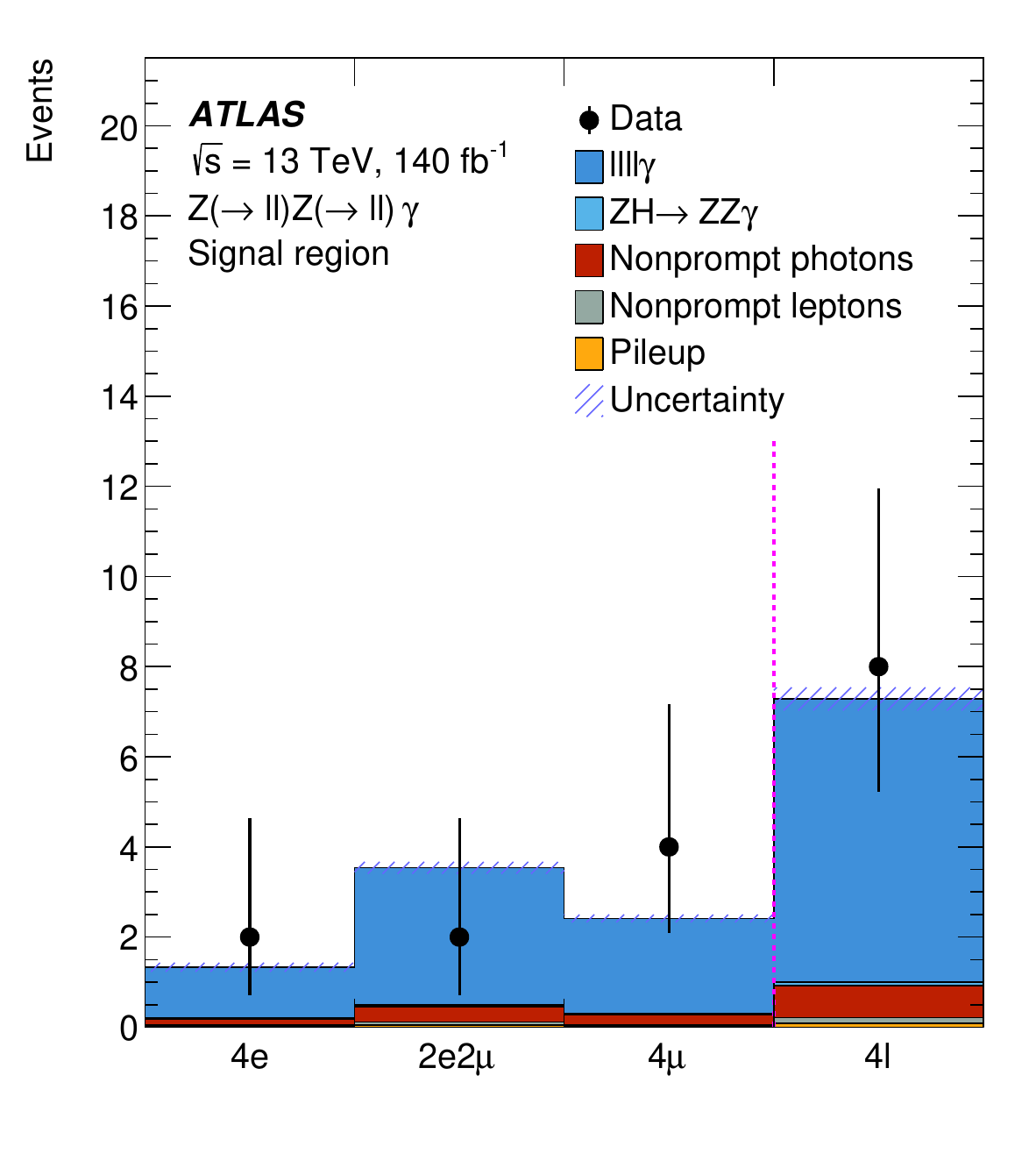}}
\subfloat[]{\includegraphics[width=0.49\textwidth]{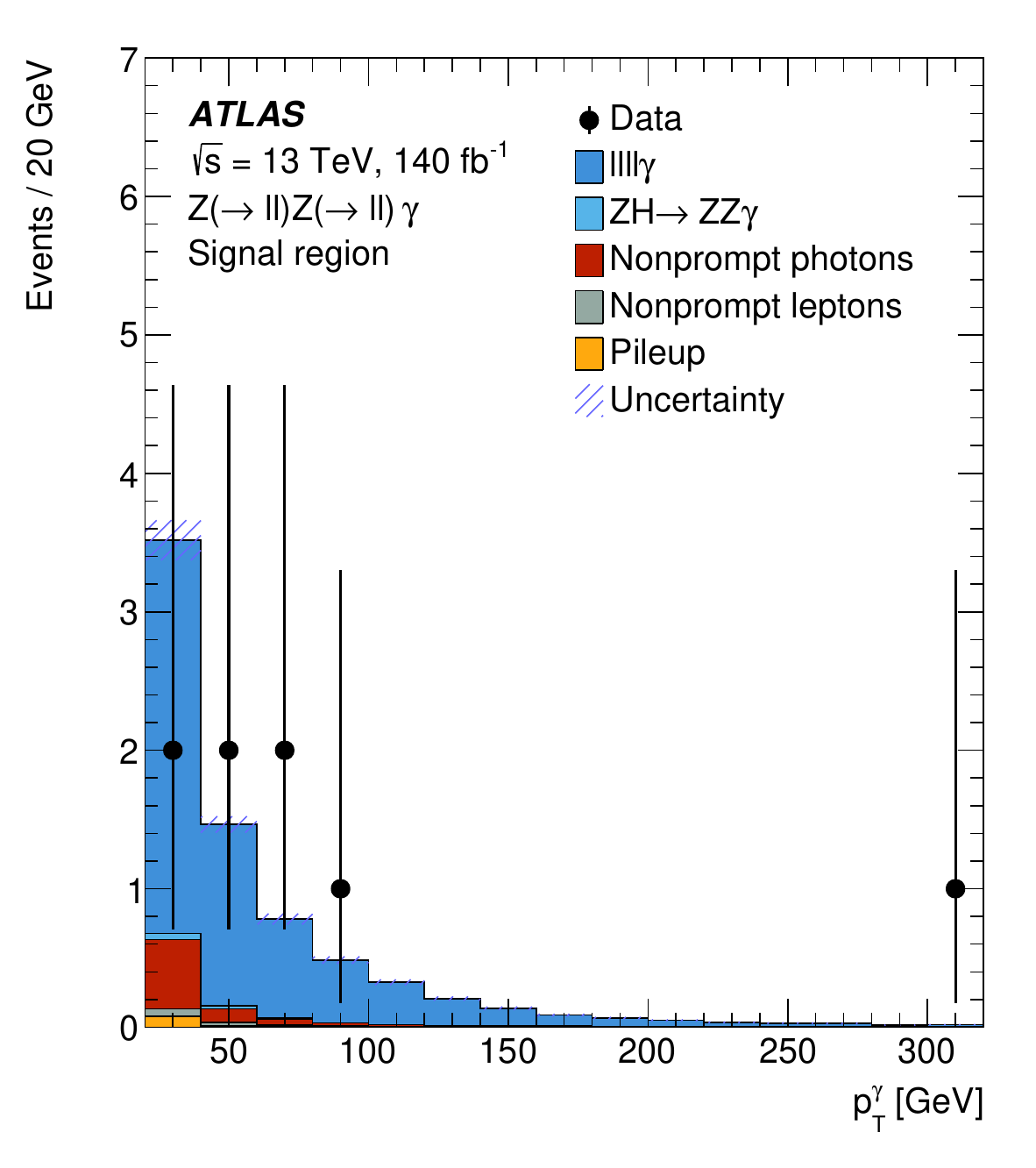}} \\
\subfloat[]{\includegraphics[width=0.49\textwidth]{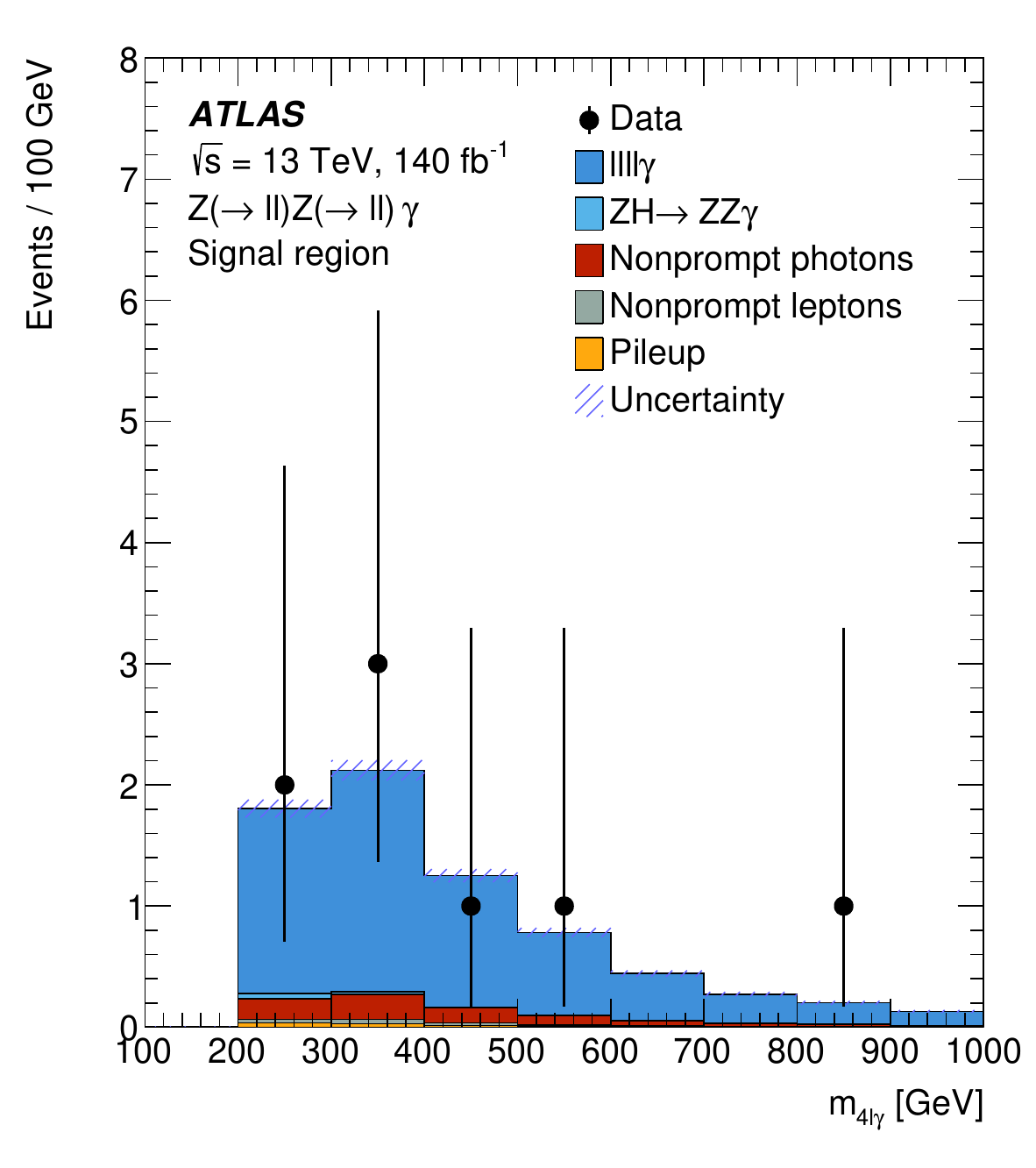}}
\subfloat[]{\includegraphics[width=0.49\textwidth]{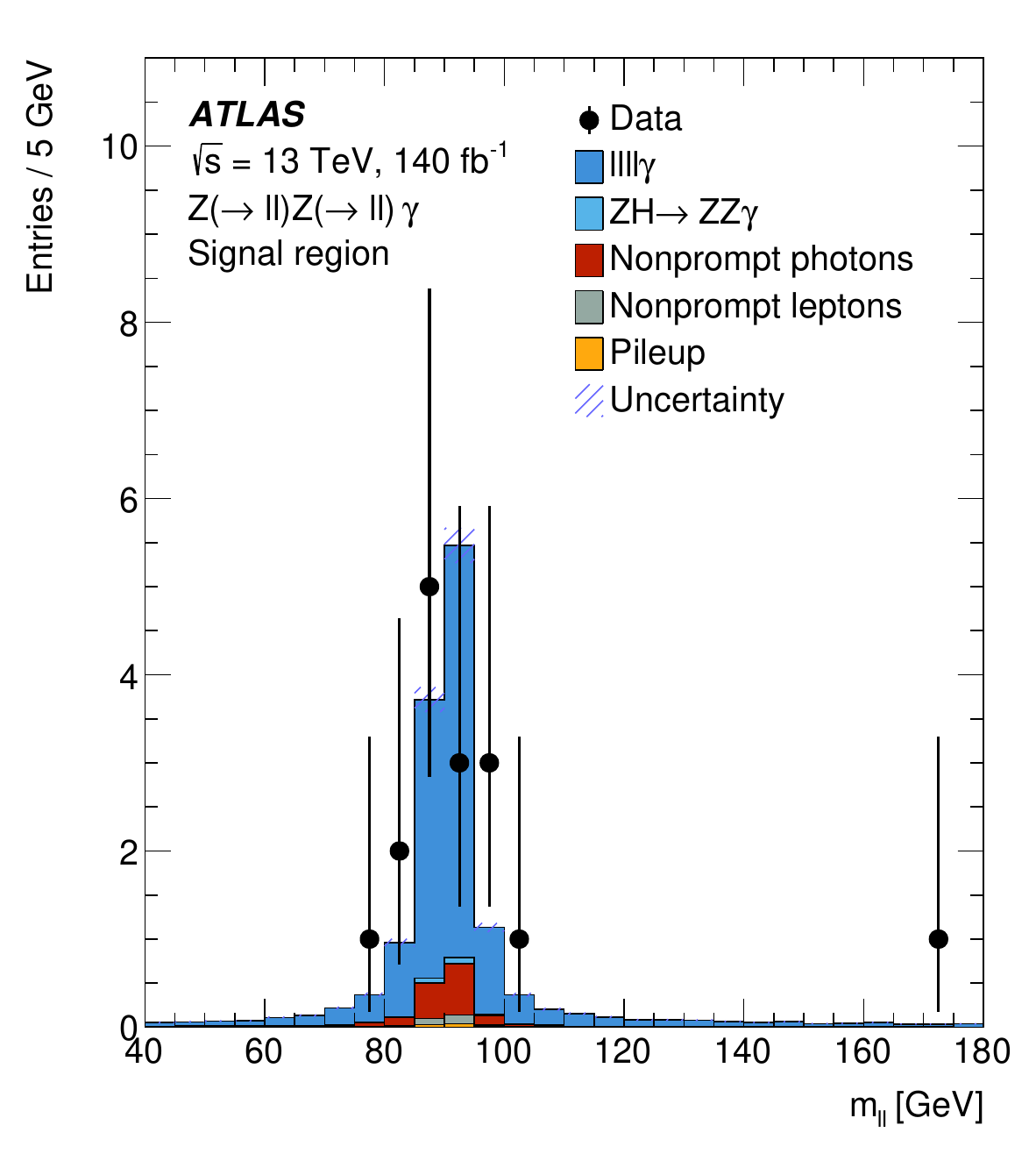}}
\caption{Data are compared with the signal and background expectations in the SR. In (a) the event yields are shown in the $4e$, $4\mu$, $2e2\mu$, and $4\ell$ channels, where the $4\ell$ channel is the sum of the other three channels, as distinguished by the dashed line. Distribution of (b) the photon transverse momentum \pTy, (c) the four-lepton-plus-photon invariant mass $m_{4\ell\gamma}$, and (d) the dilepton invariant mass $m_{\ell\ell}$, which includes entries from the leading and subleading $Z$ boson.}
\label{fig:results}
\end{figure}

A summary of the number of signal and background events is shown in~\cref{tab:overview}. Distributions in the different decay channels, the transverse momentum of the photon, the four-lepton-plus-photon invariant mass, $m_{4\ell\gamma}$, and the dilepton invariant mass of the leading and subleading $Z$ boson are displayed in~\cref{fig:results}. The analysis is dominated by statistical uncertainties. The different lepton-flavor final-state yields are consistent with expectations, and good agreement between data and prediction is observed for all distributions, with $1.0$ events ($1.3$ entries) expected in the region $p_{\mathrm{T}}^\gamma > 120\,$GeV ($m_{\ell\ell}>120\,$GeV).
The background-only hypothesis is rejected with a significance of 4.4 standard deviations, with an expected significance of 4.0 standard deviations.
The measured cross section in the fiducial region is $\sigma_{\ZZy} = \mu_{\ZZy} \sigma_{\textrm{fid.}}^{\textrm{SM}} = 0.144 _{-0.051}^{+0.065} \text{ (stat.)} _{-0.004}^{+0.008} \text{(syst.)}\,$~fb. The observed signal strength is $\mu_{\ZZy} = 1.11 _{-0.39}^{+0.50} \text{ (stat.)} _{-0.03}^{+0.05} \text{ (syst.)} _{-0.03}^{+0.06} \text{ (theory)}$ and the predicted SM fiducial cross section of \(pp\rightarrow ZZ\gamma\rightarrow\ell^+\ell^-\ell'^+\ell'^-\gamma\) is $\sigma_{\textrm{fid.}}^{\textrm{SM}} = 0.130\pm 0.001~({\text{stat.}}) \pm 0.002~({\text{PDF+}}\alpha_S) \pm0.005~({\text{scale}})~\textrm{fb}$.
This cross section includes all leptonic final states in the SR.

\begin{table}
\centering
\caption{Overview of the expected number of signal and background events for the full Run-2 dataset and the observed number of events.}
\begin{tabular}{ll}
\toprule
& {Number of events} \\[0.1cm]
\midrule
\textbf{Signal} & $6.36\pm 0.09 \ \,\text{ (stat.)} \pm  0.18 \text{ (syst.)}$ \\
\midrule
Nonprompt photons & $0.71 \pm 0.02 \  \,\text{ (stat.)} \pm 0.13 \text{ (syst.)}$ \\
Nonprompt leptons &  $ 0.12 \pm 0.02 \ \,\text{ (stat.)} \pm 0.08 \text{ (syst.)}$ \\
Pileup & $0.09 \pm 0.002\,\text{(stat.)} \pm 0.02  \text{ (syst.)}$\\[0.1cm]
\textbf{Total background} & $0.92 \pm 0.03 \ \,\text{ (stat.)} \pm 0.15 \text{ (syst.)}$ \\
\midrule
$N_\text{tot}$ prediction & $7.28 \pm 0.09 \ \,\text{ (stat.)} \pm 0.23 \text{ (syst.)}$ \\
\midrule
\textbf{$N_\text{tot}$ data} & $8$\\
\bottomrule
\end{tabular}
\label{tab:overview}
\end{table}

\FloatBarrier

The first evidence of $\ZZy$ production in $pp$ collisions is obtained using data from the ATLAS detector at the LHC. Eight events with four leptons, containing two opposite-sign same-flavor pairs, and one photon are selected from a 140~fb$^{-1}$ dataset collected in $\sqrt{s} = 13$~\TeV\ $pp$ collisions. The background-only hypothesis is rejected with an observed (expected) significance of 4.4 (4.0) standard deviations. While the event statistics in the analysis are low, the background is well controlled, and amounts to (12 $\pm$ 3)\% of the sample, illustrating the robustness of the result. The $pp\rightarrow ZZ\gamma\rightarrow\ell^+\ell^-\ell'^+\ell'^-\gamma$ cross section in the fiducial phase space, defined by kinematic requirements on the four-leptons-plus-photon system and by isolation requirements on the photon, is measured to be $\sigma_{ZZ\gamma}$ = 0.144 $_{-0.051}^{+0.065}$ (stat.) $_{-0.004}^{+0.008}$ (syst.)\,fb and is in agreement with the SM prediction.

Note added---The CMS Collaboration recently presented a similar analysis~\cite{CMS-SMP-24-014}.
\section*{Acknowledgements}
%

%

%
%

%
%

We thank CERN for the very successful operation of the LHC and its injectors, as well as the support staff at
CERN and at our institutions worldwide without whom ATLAS could not be operated efficiently.

The crucial computing support from all WLCG partners is acknowledged gratefully, in particular from CERN, the ATLAS Tier-1 facilities at TRIUMF/SFU (Canada), NDGF (Denmark, Norway, Sweden), CC-IN2P3 (France), KIT/GridKA (Germany), INFN-CNAF (Italy), NL-T1 (Netherlands), PIC (Spain), RAL (UK) and BNL (USA), the Tier-2 facilities worldwide and large non-WLCG resource providers. Major contributors of computing resources are listed in Ref.~\cite{ATL-SOFT-PUB-2026-001}.

We gratefully acknowledge the support of ANPCyT, Argentina; YerPhI, Armenia; ARC, Australia; BMWFW and FWF, Austria; ANAS, Azerbaijan; CNPq and FAPESP, Brazil; NSERC, NRC and CFI, Canada; CERN; ANID, Chile; CAS, MOST, NSFC and FRFCU, China; Minciencias, Colombia; MEYS CR, Czech Republic; DNRF and DNSRC, Denmark; IN2P3-CNRS and CEA-DRF/IRFU, France; SRNSFG, Georgia; BMFTR, HGF and MPG, Germany; GSRI, Greece; RGC and Hong Kong SAR, China; ICHEP and Academy of Sciences and Humanities, Israel; INFN, Italy; MEXT and JSPS, Japan; CNRST, Morocco; NWO, Netherlands; RCN, Norway; MNiSW, Poland; FCT, Portugal; MNE/IFA, Romania; MSTDI, Serbia; MSSR, Slovakia; ARIS and MVZI, Slovenia; DSI/NRF, South Africa; MICIU/AEI, Spain; SRC and Wallenberg Foundation, Sweden; SERI, SNSF and Cantons of Bern and Geneva, Switzerland; NSTC, Taipei; TENMAK, T\"urkiye; STFC/UKRI, United Kingdom; DOE and NSF, United States of America.

Individual groups and members have received support from BCKDF, CANARIE, CRC and DRAC, Canada; CERN-CZ, FORTE and PRIMUS, Czech Republic; COST, ERC, ERDF, Horizon 2020 and Marie Sk{\l}odowska-Curie Actions, European Union; Investissements d'Avenir Labex, Investissements d'Avenir Idex and ANR, France; DFG and AvH Foundation, Germany; Herakleitos, Thales and Aristeia programmes co-financed by EU-ESF and the Greek NSRF, Greece; BSF-NSF and MINERVA, Israel; NCN and NAWA, Poland; La Caixa Banking Foundation, CERCA and AGAUR programs from Generalitat de Catalunya and PROMETEO and GenT Programmes Generalitat Valenciana, Spain; G\"{o}ran Gustafssons Stiftelse, Sweden; The Royal Society and Leverhulme Trust, United Kingdom; Eric and Wendy Schmidt Fund for Strategic Innovation, United States of America.

In addition, individual members wish to acknowledge support from Chile: Agencia Nacional de Investigaci\'on y Desarrollo (ANID FONDECYT reg. 1230987, FONDECYT 1230812, FONDECYT 1240864, Fondecyt 3240661, Fondecyt Regular 1240721); China: Fundamental Research Funds for the Central Universities (010-63263105), Chinese Ministry of Science and Technology (MOST-2023YFA1605700, MOST-2023YFA1609300), National Natural Science Foundation of China (NSFC 12275265, NSFC-W2543005); Czech Republic: Czech Science Foundation (GACR - 24-11373S), Ministry of Education Youth and Sports (ERC-CZ-LL2327, FORTE CZ.02.01.01/00/22\_008/0004632), PRIMUS Research Programme (PRIMUS/21/SCI/017); EU: H2020 European Research Council (ERC - 101002463); European Union: European Research Council (BARD No. 101116429, ERC - 101219398, ERC - 948254, ERC 101089007), European Regional Development Fund (HE COFUND GA No.101081355, ERDF), Marie Sklodowska-Curie Actions (GAP-101168829); France: Agence Nationale de la Recherche (ANR-21-CE31-0013, ANR-22-EDIR-0002, ANR-24-CE31-0504-01); Germany: Deutsche Forschungsgemeinschaft (DFG - 469666862); China: Research Grants Council (GRF); Italy: Ministero dell'Università e della Ricerca (NextGenEU 153D23001490006 M4C2.1.1, NextGenEU I53D23000820006 M4C2.1.1, NextGenEU I53D23001490006 M4C2.1.1, SOE2024\_0000023); Japan: Japan Society for the Promotion of Science (JSPS KAKENHI  JP25H0063, JSPS KAKENHI JP22H01227, JSPS KAKENHI JP22H04944, JSPS KAKENHI JP22KK0227, JSPS KAKENHI JP24K23939, JSPS KAKENHI JP24KK0251, JSPS KAKENHI JP25H00650, JSPS KAKENHI JP25H01291, JSPS KAKENHI JP25K01011, JSPS KAKENHI JP25K01023); Norway: Research Council of Norway (RCN-314472); Poland: Polish National Science Centre (NCN 2021/42/E/ST2/00350, NCN OPUS 2023/51/B/ST2/02507, NCN OPUS nr 2022/47/B/ST2/03059, NCN UMO-2019/34/E/ST2/00393, UMO-2022/47/O/ST2/00148, UMO-2023/49/B/ST2/04085, UMO-2023/51/B/ST2/00920, UMO-2024/53/N/ST2/00869); Spain: Agència de Gestió d'Ajuts Universitaris i de Recerca. (AGAUR - 2023 BP 00141), Ministry of Science and Innovation (RYC2019-028510-I, RYC2020-030254-I, RYC2021-031273-I, RYC2022-038164-I), Ministerio de Ciencia, Innovación y Universidades/Agencia Estatal de Investigaci\'on (EU NextGenerationEU (PRTR-C17.I1), PID2022-142604OB-C22); Sweden: Carl Trygger Foundation (Carl Trygger Foundation CTS 22:2312), Swedish Research Council (Swedish Research Council 2023-04654, VR 2021-03651, VR 2022-03845, VR 2022-04683, VR 2023-03403, VR 2024-05451, VR 2025-05940), Knut and Alice Wallenberg Foundation (KAW 2023.0366); Switzerland: Swiss National Science Foundation (SNSF - PCEFP2\_194658); United Kingdom: The Binks Trust, Royal Society (NIF-R1-231091); United States of America: U.S. Department of Energy (ECA DE-AC02-76SF00515), John Templeton Foundation (John Templeton Foundation 63206), Neubauer Family Foundation.

%
%


%
\clearpage
\appendix
\FloatBarrier
\part*{End Matter}
\addcontentsline{toc}{part}{End Matter}
An overview of the event selection in the SR is given in \cref{tab:evnt_sel}, summarizing the selection criteria described earlier for the different selected particle types.

The impact of the uncertainties on the measured cross section in the fiducial region is given in \cref{tab:syst_SRUnblind}. The table lists the systematic uncertainties separated into background uncertainties, experimental uncertainties, and others. The total systematic uncertainty is much smaller than the statistical uncertainty with this data sample.

\begin{table}[ht]
\centering
\caption{Overview of the event selection in the SR.}
\begin{tabular}{l ccc}
\toprule
& Electrons & Muons & Photon \\
\midrule
Trigger & \multicolumn{2}{c}{single and dilepton trigger} &  \\
Multiplicity & \multicolumn{2}{c}{$\geq 2$ OSSF pairs} & $\geq 1$ photon \\[0.1cm]
\midrule
\multirow{2}{*}{Kinematics} & \multicolumn{2}{c}{$p_{\mathrm{T}} > 30, 25, 10, 10\,$GeV} & $p_{\mathrm{T}} > 20\,$GeV \\[0.1cm]
& $|\eta| < 2.47$ & $|\eta| < 2.5$ & $|\eta| < 2.37$ \\
Excluded & $ 1.37 < |\eta| < 1.52$ & Not applicable & $ 1.37 < |\eta| < 1.52$ \\[0.1cm]
\midrule
\multirow{2}{*}{Isolation} & Loose~\cite{EGAM-2021-01} & PflowLoose~\cite{MUON-2018-03} & Loose~\cite{EGAM-2021-01}\\[0.1cm]
& \multicolumn{2}{c}{$\Delta R(\mu, e) > 0.2$} & $\Delta R(\ell, \gamma) > 0.4$ \\[0.1cm]
Identification & Medium~\cite{EGAM-2021-01} & Medium~\cite{MUON-2018-03} & Tight~\cite{EGAM-2021-01} \\
\midrule
Invariant mass & \multicolumn{3}{c}{$m_{\ell\ell} >40$ GeV, $(m_{\ell\ell\gamma} + m_{\ell\ell}) > 2 m_Z$} \\
\bottomrule
\end{tabular}
\label{tab:evnt_sel}
\end{table}

\begin{table}
\begin{center}
\caption{The impact of uncertainties on the measured cross section in the fiducial region, $\sigma_{\ZZy}$.}
\begin{adjustbox}{max width=\textwidth}
\begin{tabular}{lc}
\toprule
Source                            & Relative uncertainty [\%] \\
\midrule\midrule
Nonprompt photon background          &1.9  \\
Nonprompt lepton background          &1.2  \\
Pileup background                     &0.4  \\
\midrule
Photon identification and isolation efficiency  & 1.6 \\
Pileup reweighting     & 1.4  \\
Muon identification, isolation, reconstruction, momentum resolution and scale   & 0.9\\
Electron identification, isolation, reconstruction efficiency & 0.6\\
Electron-photon resolution and energy scale  & 0.5\\
\midrule
Signal PDF and $\alpha_s$, QCD scales &1.4  \\
Signal MC statistics                  &1.4  \\
Integrated luminosity                 &0.9  \\
\midrule
\textbf{Total systematic uncertainty}            &4.0 \\
\midrule\midrule
Data statistics                        &40 \\
\midrule
\textbf{Total uncertainty}                       &41 \\
\bottomrule
\end{tabular}
\end{adjustbox}
\label{tab:syst_SRUnblind}
\end{center}
\end{table}

\FloatBarrier
\printbibliography
\clearpage
 
\begin{flushleft}
\hypersetup{urlcolor=black}
{\Large The ATLAS Collaboration}

\bigskip

\AtlasOrcid[0000-0002-6665-4934]{G.~Aad}$^\textrm{\scriptsize 102}$,
\AtlasOrcid[0000-0001-7616-1554]{E.~Aakvaag}$^\textrm{\scriptsize 17}$,
\AtlasOrcid[0000-0002-5888-2734]{B.~Abbott}$^\textrm{\scriptsize 121}$,
\AtlasOrcid[0000-0002-0287-5869]{S.~Abdelhameed}$^\textrm{\scriptsize 83b}$,
\AtlasOrcid[0000-0002-1002-1652]{K.~Abeling}$^\textrm{\scriptsize 54}$,
\AtlasOrcid[0000-0001-5763-2760]{N.J.~Abicht}$^\textrm{\scriptsize 48}$,
\AtlasOrcid[0000-0002-8496-9294]{S.H.~Abidi}$^\textrm{\scriptsize 30}$,
\AtlasOrcid[0009-0003-6578-220X]{M.~Aboelela}$^\textrm{\scriptsize 44}$,
\AtlasOrcid[0000-0002-9987-2292]{A.~Aboulhorma}$^\textrm{\scriptsize 36e}$,
\AtlasOrcid[0000-0001-5329-6640]{H.~Abramowicz}$^\textrm{\scriptsize 154}$
% ...
\bigskip
\\

% ...
$^{17}$Department for Physics and Technology, University of Bergen, Bergen; Norway.\\
% ...
$^{30}$Physics Department, Brookhaven National Laboratory, Upton NY; United States of America.\\
% ...
$^{36}$$^{(a)}$Facult\'e des Sciences Ain Chock, Universit\'e Hassan II de Casablanca;$^{(b)}$Facult\'{e} des Sciences, Universit\'{e} Ibn-Tofail, K\'{e}nitra;$^{(c)}$Facult\'e des Sciences Semlalia, Universit\'e Cadi Ayyad, LPHEA-Marrakech;$^{(d)}$LPMR, Facult\'e des Sciences, Universit\'e Mohamed Premier, Oujda;$^{(e)}$Facult\'e des sciences, Universit\'e Mohammed V, Rabat;$^{(f)}$Institute of Applied Physics, Mohammed VI Polytechnic University, Ben Guerir; Morocco.\\
% ...
$^{44}$Physics Department, Southern Methodist University, Dallas TX; United States of America.\\
% ...
$^{48}$Fakult\"{a}t Physik, Technische Universit{\"a}t Dortmund, Dortmund; Germany.\\
% ...
$^{54}$II. Physikalisches Institut, Georg-August-Universit\"{a}t G\"ottingen, G\"ottingen; Germany.\\
% ...
$^{83}$$^{(a)}$Khalifa University of Science and Technology, Abu Dhabi;$^{(b)}$New York University Abu Dhabi, Abu Dhabi;$^{(c)}$United Arab Emirates University, Al Ain;$^{(d)}$University of Sharjah, Sharjah; United Arab Emirates.\\
% ...
$^{102}$CPPM, Aix-Marseille Universit\'e, CNRS/IN2P3, Marseille; France.\\
% ...
$^{121}$Homer L. Dodge Department of Physics and Astronomy, University of Oklahoma, Norman OK; United States of America.\\
% ...
$^{154}$Raymond and Beverly Sackler School of Physics and Astronomy, Tel Aviv University, Tel Aviv; Israel.\\
% ...
$^{a}$ Also at Affiliated with an institute formerly covered by a cooperation agreement with CERN.\\
$^{b}$ Also at An-Najah National University, Nablus; Palestine.\\
$^{c}$ Also at Borough of Manhattan Community College, City University of New York, New York NY; United States of America.\\
$^{d}$ Also at Center for Interdisciplinary Research and Innovation (CIRI-AUTH), Thessaloniki; Greece.\\
$^{e}$ Also at Centre of Physics of the Universities of Minho and Porto (CF-UM-UP); Portugal.\\
$^{f}$ Also at CERN, Geneva; Switzerland.\\
% ...

\end{flushleft}


%

%
%

%

%
%
%
%

\end{document}